\documentclass[12pt]{article}

\usepackage{epsf}
\usepackage[dvips]{graphicx}
\newcommand\ba{\begin{eqnarray}}
\newcommand\ea{\end{eqnarray}}

\newcommand{\be}{\begin{equation}}
\newcommand{\ee}{\end{equation}}
\begin{document}
\def\vP{{\vec {P}}}
\def\vA{{\vec {A}}}
\def\vq{{\vec {q}}}
\def\vk{{\vec {k}}}
\def\vp{{\vec {p}}}
\def\vs{{\vec {s}}}
\def\vu{{\vec {1}}}
\def\vpu{{\vec {p_1}}}
\def\vPu{{\vec {P_1}}}
\def\vpd{{\vec {p_2}}}
\def\vPd{{\vec {P_2}}}
\def\vpt{{\vec {p_3}}}
\def\vpq{{\vec {p_4}}}
\def\tq{{\vec {\widetilde{q}}}}
\def\tpu{{\vec {\widetilde{p_1}}}}
\def\tpd{{\vec {\widetilde{p_2}}}}
\def\tpt{{\vec {\widetilde{p_3}}}}
\def\tpq{{\vec {\widetilde{p_4}}}}
\def\vqb{{\vec {q_B}}}
\def\vku{{\vec {k_1}}}
\def\vkd{{\vec {k_2}}}
\def\tqb{{\vec {\widetilde{q_B}}}}
\def\tku{{\vec {\widetilde{k_1}}}}
\def\tkd{{\vec {\widetilde{k_2}}}}
\def\eu{\epsilon_{1}}
\def\ed{\epsilon_{2}}
\def\eub{\epsilon_{1B}}
\def\edb{\epsilon_{2B}}
\def\vkub{{\vec {k_{1B}}}}
\def\vkdb{{\vec {k_{2B}}}}
\def\vpub{{\vec {p_{1B}}}}
\def\vpdb{{\vec {p_{2B}}}}
\def\tkub{{\vec {\widetilde{k_{1B}}}}}
\def\tkdb{{\vec {\widetilde{k_{2B}}}}}

\begin{center}

{\bfseries From theory to experiment: hadron electromagnetic form factors in space-like and time-like regions 
}

\vskip 5mm

E. Tomasi-Gustafsson
\vskip 5mm

{\small
\it
DAPNIA/SPhN, CEA/SACLAY, 91191 Gif-sur-Yvette Cedex France 
}  

\vskip 5mm
G. I. Gakh, and A. P. Rekalo
\vskip 5mm

{\small
{\it NSC-KIPT, Akademickeskaya 1, 61108 Kharkov, Ukraine}

\vskip 5mm
\begin{minipage}{150mm}
\centerline{\bf Abstract}

Hadron electromagnetic form factors contain dynamical information on the intrinsic structure of the hadrons. The pioneering work developed at the  Kharkov Physical-Technical
Institute in the 60's on the relation between the polarized cross section and form factors triggered a number of experiments. Such experiments could be performed only recently, due to the progress in accelerator and polarimetry techniques. The principle of these measurements is recalled and the surprising and very precise results obtained on proton are presented. The actual status of nucleon electromagnetic form factors is reviewed, with special attention to the basic work done in Kharkov Institute.
\end{minipage}
}

\end{center}

\centerline{\bf Introduction}

Electromagnetic form factors (FFs) are fundamental quantities which describe the internal structure of composite particles. Hadron FFs contain dynamical information about charge and magnetic currents and are calculated in frame of hadron models. Elastic hadron FFs can be studied through  elastic electron hadron scattering $e+h\to e+h$, or through annihilation reactions 
$\bar p+p\leftrightarrow e^+ + e^-$, where the momentum is transferred by the exchange of one photon. Assuming this reaction mechanism, FFs enter in the expresssion of hadron electromagnetic vertex, and can be directly accessible from experiment, measuring the differential cross section and polarization observables. 

Polarization observables, indeed, is the key word of this talk, which is dedicated to the fundamental contribution of the 'Kharkov school', leaded by Academician A.I. Akhiezer, whose memory we honour today. Basic papers, in collaboration with Prof. M. P. Rekalo, were written in the late 60's, which indicated the way to get precise data on FFs at large values of the momentum transfer squared, $Q^2=-q^2$ \cite{Ak68,Ak74}.  Such experiments have been realized only recently, due to the progress achieved in building high intensity polarized beams, spectrometers, hadron polarimeters in the GeV range. The model independent derivation of the necessary observables, the ideas and the suggestions made in Kharkov almost 40 years ago, represent a tremendous advance of the theory on experiment. At that time it was difficult to conceive that an intense high polarized beam  could be accelerated, and the calculations were done for polarized target, which seemed more realistic.

Nowadays  higher transfer momenta are reached with polarized beam and hadron polarimeters which can measure the polarization of the scattered hadron, proton \cite{Jo00} or deuteron \cite{t20}, although if polarized targets are currently used.

In this presentation, we briefly present the main lines of the theoretical background, describe the experimental set up and focus on the results and their implications.

\section{Electron-hadron elastic scattering-Theoretical framework}
\subsection{Theoretical framework}

\begin{center}
\begin{figure}[h]
\mbox{\epsfxsize=7.cm\leavevmode\epsffile{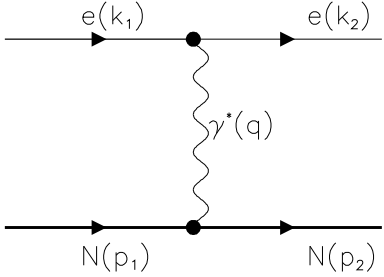}}
\caption{ Diagram for elastic scattering $e+N\to e+N$}
\label{fig:epel}
\end{figure}
\end{center}

The Feynman diagram for elastic $eN$-scattering is shown in Fig. \ref{fig:epel}, assuming one-photon exchange, together with the notations of the particle four-momenta. 
The most convenient frame for the analysis of elastic $eN$-scattering is the Breit frame, which is defined as the system where the initial and final nucleon energies are the same. As a consequence, the energy of the virtual photon vanishes and its four-momentum square, coincides with its three-momentum square (in modulus). Therefore, the derivation of the formalism in Breit system is more simple and 
has some analogy with a non-relativistic description of the nucleon electromagnetic structure. We choose the $z$-axis parallel to the photon three-momentum in the Breit system and the $xz$-plane as the scattering plane.
An useful kinemtical relation can be derived between the electron scattering angle in the Lab system $\theta_e$ and in the Breit system $\theta_B$:
\be
\cot^2\displaystyle\frac{\theta_B}{2}=\displaystyle\frac{\cot^2\theta_e/2}{1+\tau},
~\tau=\displaystyle\frac{-q^2}{4m^2}, 
\label{eq:cot}
\ee
where $m$ is the nucleon mass and $q=k_1-k_2$. The matrix element corresponding to the diagram of Fig.  \ref{fig:epel} is:
\begin{equation}
{\cal M}=\displaystyle\frac{e^2}{q^2} 
\ell_{\mu}{\cal J}_{\mu}=\displaystyle\frac{e^2}{q^2} 
\ell\cdot {\cal J},
\label{eq:mat}
\end{equation}
where $\ell_{\mu}=\overline{u}(k_2)\gamma_{\mu}u(k_1)$ is the electromagnetic current of electron. The nucleon electromagnetic current, ${\cal J}_{\mu}$,  describes the proton vertex and can be written in terms of Pauli and Dirac FFs $F_1$ and $F_2$:
\begin{equation}
{\cal J}_{\mu}=\overline{u}(p_2)\left [F_1(q^2)\gamma_{\mu}-
\displaystyle\frac{\sigma_{\mu\nu}q_\nu}{2m}F_2(q^2)\right ]u(p_1),~\sigma_{\mu\nu}=\displaystyle\frac{\gamma_{\mu}\gamma_{\nu}-
\gamma_{\nu}\gamma_{\mu}}{2}.
\label{eq:eqj}
\end{equation}
Note that ${\cal J}\cdot q=0$, for any values of $F_1$ and $F_2$, i.e.,  the current ${\cal J}_{\mu}$ is conserved. 
The expressions for the different components of the current ${\cal J}_{\mu}$ (valid in the Breit frame only) are:
$$
\begin{array}{ll}
{\cal J}_0&= 
 2m\chi_2^\dagger\chi_1\left (F_1-\tau F_2\right ),\\
\vec {\cal J}&=
i\chi_2^\dagger\vec\sigma\times\vqb \chi_1\left (F_1+F_2\right ), 
\end{array}
$$
and allow to introduce in a straightforward way the Sachs nucleon electromagnetic FFs, electric and magnetic, which are written as:
$$G_E=F_1-\tau F_2, ~~G_M=F_1+F_2.$$
Such identification can be easily understood, if one takes into account that the time component of the current, ${\cal J}_0$, describes the interaction of the nucleon with Coulomb potential. Correspondingly, the space component 
$\vec {\cal J}$ describes the interaction with the magnetic field. 

From Eq. (\ref{eq:mat}) we can find the following representation for $ \left | {\cal M}\right |^2$
$$
\left |{\cal M}\right |^2 =\left (\displaystyle\frac{e^2}{q^2}
\right )^2 \left |\ell\cdot {\cal J} \right|^2 =
\left (\displaystyle\frac{e^2}{q^2} \right )^2L_{\mu\nu}W_{\mu\nu},
$$
where $L_{\mu\nu}=\ell_{\mu}\ell_{\nu}^* $ is the leptonic tensor and $W_{\mu\nu}={\cal J}_{\mu}{\cal J}_{\nu}^* $ is the hadronic tensor.

The product of the tensors  $L_{\mu\nu}$ and $W_{\mu\nu}$ is a relativistic invariant, therefore it can be calculated in any reference system.  The differential cross section, in any coordinate system, can be expressed in terms of the matrix element as:
\begin{equation}
d\sigma=\displaystyle\frac{(2\pi)^4\left|{\cal M}\right |^2}{4\sqrt{(k_1\cdot p_1)^2-m_e^2m^2}}\delta^4(k_1+p_1-k_2-p_2)
\displaystyle\frac{d^3\vkd}{(2\pi)^3 2\epsilon_2}
\displaystyle\frac{d^3\vpd}{(2\pi)^3 2E_2},
\label{eq:csma}
\end{equation}
where $m_e$ is the electron mass and $\epsilon_2(E_2)$ is the energy of the final electron (nucleon). For comparison with experiment it is more convenient to use the differential cross section in Lab system,
${d\sigma}/{d\Omega}_e$, where $d\Omega_e$ is the element of the electron solid angle in the Lab system:
\begin{equation}
\displaystyle\frac{d\sigma}{d\Omega}_e=\displaystyle\frac
{\left| {\cal M}\right |^2}{64\pi^2}
\left (\displaystyle\frac{\ed}{\eu}\right )^2
\displaystyle\frac{1}{m^2}, 
\label{eq:csm2}
\end{equation}
where $\epsilon_1$ is the energy of the initial electron. After straightforward calculation one recovers the Rosenbluth formula \cite{Ro50}:
\begin{equation}
\displaystyle\frac{d\sigma}{d\Omega}_e=\sigma_M\left [2\tau G_M^2\tan^2\frac{\theta_e}{2} +
\displaystyle\frac{G_E^2+\tau G_M^2}{1+\tau}\right ],
\label{eq:csst}
\end{equation}
with 
$$\sigma_M=
\displaystyle\frac{4\alpha^2}{(-q^2)^2}\displaystyle\frac{\ed^3}{\eu}\cos^2 \frac{\theta_e}{2}= \displaystyle\frac{4\alpha^2}{(-q^2)^2}
\displaystyle\frac{\ed^2\cos^2\frac{\theta_e}{2}}
{1+2\displaystyle\frac{\eu}{m}\sin^2
\displaystyle\frac{\theta_e}{2}},
$$
where $\sigma_M$ is the Mott cross section, describing the scattering of unpolarized electrons by a point charge particle (with spin 1/2).

Note that the very specific $\cot^2\displaystyle\frac{\theta_e}{2}$-dependence of the reduced cross section for $eN$-scattering results from the assumption of one-photon mechanism for the considered reaction.

This can be easily proved \cite{Re99}, by cross-symmetry considerations, looking to the annihilation channel, $e^++e^-\to p+\overline{p}$. In the CMS of such reaction, the one-photon mechanism induces a simple and evident $\cos^2\theta$-dependence of the corresponding differential cross section ($\theta$ is the angle of the emitted nucleon in center of mass system), due to the C-invariance of the hadron electromagnetic interaction, and unit value of the photon spin. The particular $\cot^2\displaystyle\frac{\theta_e}{2}$-dependence of the differential $eN$-cross section is at the basis of the method to determine both nucleon electromagnetic FFs, $G_E$ and $G_M$, using the linearity of the {\it reduced} cross section:
\begin{equation}
\sigma_{red}^{Born}(\theta_e,Q^2)=\epsilon(1+\tau)\left [1+2\displaystyle\frac{\epsilon_1}{m}\sin^2(\theta_e/2)\right ]\displaystyle\frac
{4 \epsilon_1^2\sin^4(\theta_e/2)}{\alpha^2\cos^2(\theta_e/2)}\displaystyle\frac{d\sigma}{d\Omega}_e=\tau G_M^2(Q^2)+\epsilon G_E^2(Q^2),
\label{eq:sigma}
\end{equation}
$$
\epsilon=[1+2(1+\tau)\tan^2(\theta_e/2)]^{-1},
$$
where $\alpha=1/137$.  Measurements of the elastic differential cross section as a function of $\epsilon$, at different angles for a fixed value of $Q^2$ allow $G_E(Q^2)$ and $G_M(Q^2)$ to be determined as the slope and the intercept, respectively, from the linear $\epsilon$ dependence (\ref{eq:sigma}) (Rosenbluth fit) \cite{Ro50}. One can see that the backward $eN$-scattering ($\theta_e=\pi,\cot^2\frac{\theta_e}{2}=0$) is determined by the magnetic FF only, and that the slope for $\sigma_{red}$ is sensitive to $G_E$.

At large $q^2$, (such that $\tau\gg 1$), the differential cross-section ${d\sigma}/{d\Omega}_e$ (with unpolarized particles) is unsensitive to $G_E$: the corresponding combination of the nucleon FFs, $G_E^2+\tau G_M^2$ is dominated by the $G_M$ contribution, due to the following reasons:
\begin{itemize}
\item $G_{Mp}(q^2)/G_{Ep}(q^2)\le \mu_p$, where $\mu_p$ is the proton magnetic moment, so $G_{Mp}^2(q^2)/G_{Ep}^2(q^2)\le 2.79^2\simeq 8$;
\item The factor $\tau$ increases the $G_M^2$ contribution at large momentum transfer, where $\tau\gg 1$.
\end{itemize}
Therefore, $ep-$scattering (with unpolarized particles) is dominated by the magnetic FF, at large values of momentum transfer. The same holds for $en-$scattering, even at relatively small values of $q^2$, due to the smaller values of the neutron electric FF. 

As a result, for the exact determination of the proton electric FF, in the region of large momentum transfer,  and for the neutron electric FF - at any value of $q^2$, polarization measurements are required and in particular those polarization observables which are determined by the product $G_EG_M$, and are, therefore, more sensitive to $G_E$. 
Both experiments (with polarized electron beam) have been realized: $p(\vec e, \vec p) e$ for the determination of $G_{Ep}$ \cite{Jo00} and, for the determination of $G_{En}$, $\vec d(\vec e, e'n) p$  \cite{Wa03} and $d(\vec e, e'\vec n) p$ \cite{Pl05}.

In general the hadronic tensor $W_{\mu\nu}$, for $ep$ elastic scattering, contains four terms, related to the 4 possibilities of polarizing the initial and final protons:
$$W_{\mu\nu}=W_{\mu\nu}{(0)}+W_{\mu\nu}(\vPu)+W_{\mu\nu}(\vPd)+W_{\mu\nu}(\vPu,\vPd),$$
where $\vPu~(\vPd)$ is the polarization vector of the initial (final) proton.
The first term corresponds to the unpolarized case, the second (third) term corresponds to the case when the initial (final) proton is polarized, and the last term describes the reaction when both protons (initial and final) are polarized. 

One can show  that the polarization of the final proton vanishes, if the electron is unpolarized: unpolarized electrons can not induce polarization of the scattered proton. This is a property of the one-photon mechanism for elastic $eh$-scattering and of the hermiticity of the Hamiltonian for the hadron electromagnetic interaction. Namely the hermiticity condition allows to prove that the hadron electromagnetic FFs are real functions of the momentum transfer squared in the space-like region. On the other hand, in the time-like region, which is scanned by the annihilation process $e^-+e^+\leftrightarrow p+\overline{p}$, the nucleon electromagnetic FFs are complex functions of $q^2$, if $q^2\ge 4m_\pi^2$, where $m_\pi$ is the pion mass. The complexity of nucleon FF's (in the time-like region) results in specific polarization phenomena, for the annihilation process $e+e^-\leftrightarrow p+\overline{p}$, which are different from the case of elastic $ep-$scattering. For example, the polarization of the final proton (or antiproton) is different from zero, even in the case of collisions of unpolarized leptons: this polarization is determined by the product ${\cal I}m G_EG_M^*$ (and, therefore, vanishes in the case of elastic $ep$-scattering, where FFs are real). 

Note that two-photon exchange in $ep$-elastic scattering is also generating complex amplitudes. So, the interference between one- and two-photon amplitudes induces nonzero proton polarization, but small in absolute value, as it is proportional to $\alpha$.

Numerous experiments have been done with the aim to detect such polarization at small momentum transfer $|q^2|\le$ 1 GeV$^2$, but with negative result, at a percent level.
Only recently the above mentioned interference was experimentally detected, measuring the asymmetry in the scattering of transversally polarized electrons by an unpolarized proton target \cite{Ma04}, which contains information on the imaginary part of the two--photon contribution.

Note that at very large momentum transfer, the relative role of two-photon amplitudes may be increased (violating the counting in $\alpha$), due to the steep $q^2$-decreasing of hadronic electromagnetic FFs.

Note also that the analytical properties of the nucleon FFs, considered as functions of the complex variable $z=q^2$, result in a specific asymptotic behavior, as they obey to the Phragm\`en-Lindel\"of theorem:
\begin{equation}
\lim_{q^2\to -\infty} F^{(SL)}(q^2) =\lim_{q^2\to \infty} 
F^{(TL)}(q^2).
\label{eq:eqpl}
\end{equation}
The existing experimental data about the proton FFs in time-like region up to 15 GeV$^2$, seem to contradict this theorem \cite{Re01}, showing that the asymptotic region is more far than expected. 

Let us define a coordinate system where $z$ is parallel to the photon three-momentum and $xz$ is the scattering plane. One can find the following expressions for the components 
$P_x$ and $P_z$ of the proton polarization vector (in the scattering plane) - in terms of the proton electromagnetic FFs:
\begin{equation}
\begin{array}{ll}
DP_x&=-2\lambda \cot \displaystyle\frac{\theta_e}{2} \sqrt{\displaystyle\frac{\tau}{1+\tau }}G_EG_M,\\
DP_z&=\lambda\displaystyle\frac{\eu+\ed}{m}\sqrt{\displaystyle\frac{\tau}{1+\tau }}G_M^2,\\
\end{array}
\label{eq:fi}
\end{equation}
where $\lambda$ is the electron helicity, which takes values $\pm 1$, corresponding to the direction of spin parallel or antiparallel to the electron three-momentum, and $D$ is proportional to the differential cross section with unpolarized particles:
\begin{equation}
D=2\tau G_M^2+\cot^2 \displaystyle\frac{\theta_e}{2} \displaystyle\frac{G_E^2+\tau G_M^2}{1+\tau }.
\label{eq:mott}
\end{equation}
So, for the ratio of these components one can find the following formula:
\begin{equation}
\displaystyle\frac{P_x}{P_z}=\displaystyle\frac{P_t}{P_\ell}= - 2\cot \displaystyle\frac{\theta_e}{2} \displaystyle\frac{m}{\eu+\ed}\displaystyle\frac{G_E(q^2)}{G_M(q^2)}.
\label{eq:final}
\end{equation}
A  measurement of the ratio of the transverse and the longitudinal polarization of the recoil proton is directly related to the ratio of electric and magnetic FFs, $G_E(q^2)/G_M(q^2)$.
In the same way it is possible to calculate the dependence of the differential cross section for the elastic scattering of the longitudinally polarized electrons by a {\bf polarized} proton target, with polarization ${\cal P}$:
\begin{equation}
\displaystyle\frac{d\sigma}{d\Omega_e}({\cal P})=
\left (\displaystyle\frac{d\sigma}{d\Omega_e}\right )_0 
\left ( 1+\lambda {\cal P}_xA_x+\lambda {\cal P}_zA_z\right ),
\label{eq:tpol}
\end{equation}
where the asymmetries $A_x$ and $A_z$ (or the corresponding analyzing powers) are related in a simple and direct way, to the components of the final proton polarization:
\begin{equation}
\begin{array}{ll}
A_x&=P_x,\\
A_z&=-P_z.\\
\end{array}
\label{eq:ap}
\end{equation}
This holds in the framework of the one-photon mechanism for elastic $ep-$scattering. Note that the quantities 
$A_x$ and $P_x$ have the same sign and absolute value, but the components $A_z$ and $P_z$, being equal in absolute value, have opposite sign.

In the framework of the one-photon mechanism, there are at least two different sources of corrections to these relations:
\begin{itemize}
\item the standard radiative corrections;
\item the electroweak corrrections.
\end{itemize}
\subsection{Experimental results}

Highly polarized electron beams are available at different accelerators, MAMI (Mainz), MIT (Bates), JLab (Virginia). At JLab, where the $G_{Ep}$ experiment was done \cite{Jo00}, energies are  available up to 6 GeV, the typical intensity about 30 $\mu$ A and the polarization from 60 to 80\%. The (longitudinal) polarization is obtained by photoemission from a semiconductor cathode using polarized laser light from a pulsed diode laser. Beam polarimeters based on Mott, Moeller or Compton scattering measure the beam polarization with an error of the order of percent.  

Measurements of elastic $ep$ scattering require coincidence experiments in order to eliminate the background, even if in a binary process, in principle, the detection of one particle allows to fully reconstruct the kinematics. 

The momentum of the scattered proton is analyzed by a high resolution spectrometer which focal plane detection constitutes also the front detection of the focal polarimeter.

Proton polarimeters in the GeV range are based on inclusive scattering on a graphite or polyethylene target, where one charged particle is detected. The azymuthal asymmetry of the scattered particle contain the information on the polarization of the proton at the focal plane. 

The optimization of the figure of merit (efficiency and analyzing powers) in the GeV range was carefully studied at Saturne accelerator \cite{Pomme} and at JINR-LHE accelerator complex in Dubna, where polarized proton beams are available in the GeV range \cite{Sitnik}. In particular it has been shown that the thickness of the target is a very important parameter, and depends on the proton energy. For proton momenta over 3 GeV/c, very thick targets (larger than the collision length) do not improve the polarimeter performances.

\begin{center}
\begin{figure}
\mbox{\epsfxsize=10.cm\leavevmode \epsffile{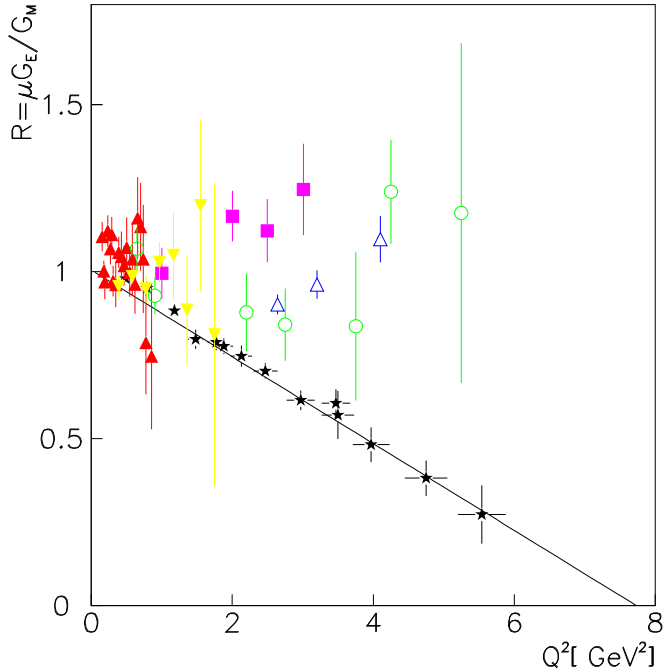}}
\caption{$Q^2$-dependence for the proton  form factor ratio. Selected data from Rosenbluth measurements are plotted: from Ref. \protect\cite{Ja66} (solid triangles); from Ref. \protect\cite{An94} (solid circles); from Ref. \protect\cite{Ch04} (open circles); from Ref. \protect\cite{Wa94} (solid squares); 
from Ref. \protect\cite{Qa05} (open triangles). Polarization data (solid stars) are shown together with the fit from Eq. (\protect\ref{eq:brash}).}
\label{fig:gegm}
\end{figure}
\end{center}

After that a good elastic $ep$ event is identified by energy and angular correlation of the two outgoing particles, and its polarization measured, the ratio of the longitudinal and transverse polarization is directly related to the ratio $R=\mu G_E/G_M$ ($\mu$ is the proton magnetic moment), by Eq. (\ref{eq:final}). The ratio $R=\mu G_E/G_M$ is shown in Fig. \ref{fig:gegm}. These data show two remarkable features: high precision of the polarization data, compared to the Rosenbluth data and  monotonical decrease  with $Q^2$ which can be parametrized as:
\begin{equation}
R(Q^2)=1-(0.130\pm 0.005)\{Q^2~[\mbox{GeV}^2]-(0.04\pm 0.09)\}.
\label{eq:brash}
\end{equation}

\subsection{Implications}

The recent polarization data show that the charge and magnetic currents in the nucleon are different, contrary to what had been previously assumed. Indeed, a commonly used  parametrization was a dipole approximation for both FFs, which was compatible with an exponential distribution of the charge, in non relativistic approximation, and also agreed with predictions from quark counting rules \cite{Ma73}. 

QCD predicts a  $Q^2F_2/F_1$ scaling, because $F_2$ carries an extra factor $1/Q^2$ as it requires a spin flip. Such scaling was approximately in agreement with the previous data, and it was argued that asymptotic predictions were reached already at $Q^2\sim 2\div$ 5 GeV$^2$. When  logarithmic corrections are added, pQCD parametrizations may reproduce the polarization data (which scale as  $Q F_2/F_1$). However analyitical properties of FFs, which should satisfy Phragm\`en-Lindeloff theorem, are fulfilled by such parametrizations only at much larger value of $Q^2$ \cite{ETG05}.

Another important issue concerns the light nuclei structure, $d$, $^3\!He$. A good description of the electromagnetic properties of these nuclei requires the knowledge of nucleon FFs. A modification of the proton FFs requires either another prescription for the neutron FFs, or a revision of other ingredients of the models, such as wave functions or meson exchange currents, relativistic effects etc \cite{ETG01}.

Data in time-like region are also necessary for a complete understanding of the nucleon structure. The separation of individual FFs has not been done, yet. As FFs are complex in TL region, polarization experiments are necessary. The present understanding is poor, and few phenomenological models can describe all data in the full kinematical region \cite{ETGtl}. Experiments are planned in future, at Novosibirsk, Frascati, FAIR.

Reasons of the discrepancy between the two methods have been indicated in two photon exchange \cite{Af05}. However this is incompatible with model independent considerations, which require non-linearity of the Rosenbluth fit as a function of $\epsilon $ and the experiments do not give evidence for the presence of such mechanism \cite{Re03,ETG04}. Recent calculations of the box diagram prove that this contribution is small \cite{Ku06,Bo06}. A more realistic explanation relies on the method used to calculate and to apply standard radiative corrections, as a multiplicative factor to the measured cross section. Such factor contains a large $\epsilon$ and $Q^2$ dependence, which are the relevant variables. Therefore, this procedure induces large correlations between the parameters of the Rosenbluth fit \cite{ETG06}. A recent suggestion to apply higher order corrections, through the structure function method, proves to be successful in bringing into agreement the two sets of data \cite{By06}.

\section{Conclusions}

The pioneering work started in Kharkov is at the origin of a series of experiments and programs at different world accelerators. The unexpected results which were obtained changed our view on the nucleon structure. 

Although an english translation of the original papers \cite{Ak68} was soon available, these papers did not receive the consideration they deserve and are not properly quoted. Only recently these papers have been added to the High-Energy Physics Literature Data base \cite{spires} and appear very little quoted in comparison with later works. From this data basis, it appears today that works, which essentially reproduce the same result are more quoted, as  for example  Ref. \cite{Do69} which appeared at least one year later and  Ref. \cite{Ar80} published in 1981.

This situation is not new, and unfortunately not adequate citation of appropriate references is a current problem. Efforts and interventions of scientific authorities at the level of Editoral journals, Conference Committees, representatives and individuals are necessary in order to have a proper recognizements of these achievements.

We would like to conclude with a citation from Ref. \cite{Ak68}: 
'{\it Thus, there exist a number of polarization experiments which are more effective for determining the proton charge form factor than is the measurement of the differential cross section for unpolarized particles}'.

\end{document}